\documentstyle[sprocl]{article}

\input{epsf}
\newcommand{\eg}{{\em e.g.\ }}

\def\be{\begin{equation}}
\def\ee{\end{equation}}
\def\bea{\begin{eqnarray}}
\def\eea{\end{eqnarray}}

\begin{document}

\title{
\begin{flushright}
{\normalsize IIT-HEP-95/5\\
hep-ex/9510002
}
\end{flushright}
\vskip 0.05in
PRODUCTION OF CHARM, CHARMONIUM, AND BEAUTY IN 800$\,$GEV
PROTON-NUCLEON COLLISIONS~\footnote{to appear in {\sl Proc.\
6th Int.\ Conf.\ on Hadron Spectroscopy (Hadron 95)},
Manchester, England, 10--14 July 1995.}
\vskip 0.05in
}

\author{
\vspace{-0.25 in}
 DANIEL M. KAPLAN~\footnote{E-Mail address: kaplan@fnal.gov.} }

\address{Illinois Institute of Technology,\\
Chicago, IL 60616, USA}

\author{
\vspace{-0.1 in}
for the Fermilab E789 Collaboration~\footnote{
C.~N.~Brown, T.~A.~Carey,
Y.~C.~Chen, R.~Childers, W.~E.~Cooper, C.~W.~Darden, G.~Gidal, K.~N.~Gounder,
P.~M.~Ho, L.~D.~Isenhower, D.~M.~Jansen,
R.~G.~Jeppesen, D.~M.~Kaplan, J.~S.~Kapustinsky, G.~C.~Kiang,
M.~S.~Kowitt, D.~W.~Lane, L.~M.~Lederman, M.~J.~Leitch, J.~W.~Lillberg,
W.~R.~Luebke, K.~B.~Luk, P.~L.~McGaughey, C.~S.~Mishra, J.~M.~Moss,
J.~C.~Peng, R.~S.~Preston, D.~Pripstein, J.~Sa, M.~E.~Sadler, R.~Schnathorst,
M.~H.~Schub, V.~Tanikella, P.~K.~Teng, J.~R.~Wilson;
Abilene Christian University,
Academia Sinica (Taiwan),
University of California at Berkeley,
University of Chicago,
Fermi National Accelerator Laboratory,
Illinois Institute of Technology,
Lawrence Berkeley Laboratory,
National Cheng Kung University,
Northern Illinois University,
University of South Carolina.
}}

\maketitle\abstracts{
}

\vspace{-0.4 in}

Fermilab Experiment 789 took data during the 1991 fixed-target run on the
production of charm, charmonium, and beauty in 800\,GeV proton-nucleon
collisions.
Several results have been published or submitted for publication; only a brief
synopsis will be given here.
The apparatus is shown schematically in Figs.~1 and 2; it is based on the
pre-existing E605/772 spectrometer, augmented for this run primarily by the
addition of two silicon-microstrip vertex telescopes (Fig.~2),
one above and one below
the beam. In most events only two
oppositely-charged particle tracks traversing the entire
spectrometer are reconstructed, one passing through
each telescope.

\begin{figure}[htb]
\vspace{-1.4 in}
\centerline{\hspace{0.318 in}\epsfysize = 2.77 in \epsffile {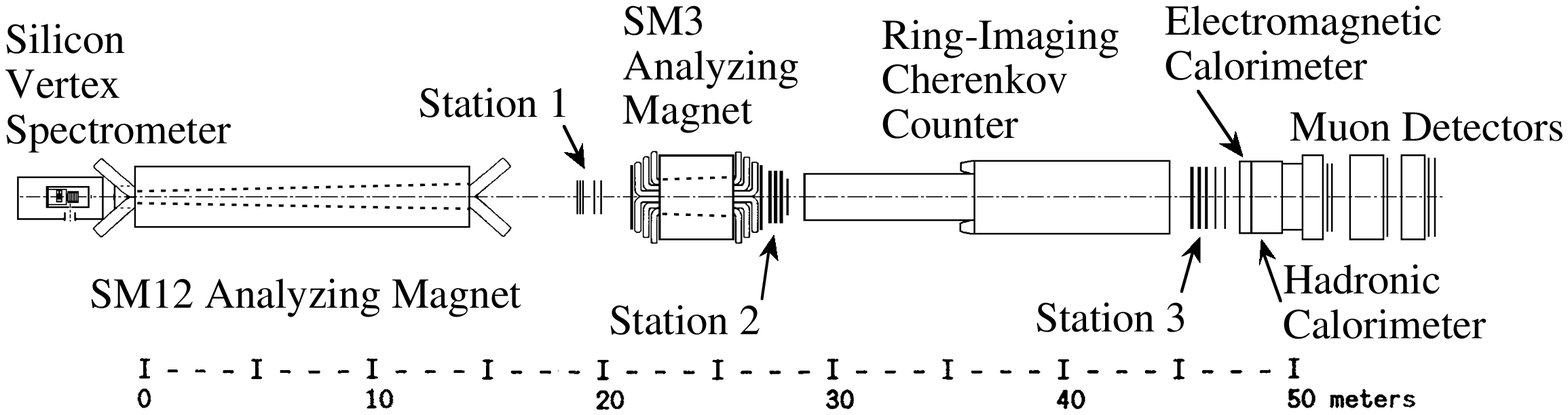}}
\vspace{-0.25 in}
\caption [E789  apparatus (plan view).]%
{E789 apparatus (plan view).}
\vspace{-0.1 in}
\end{figure}
\begin{figure}[htb]
\vspace{-.75in}
\centerline{\hspace{0.3391 in}\epsfysize = 2.4 in \epsffile {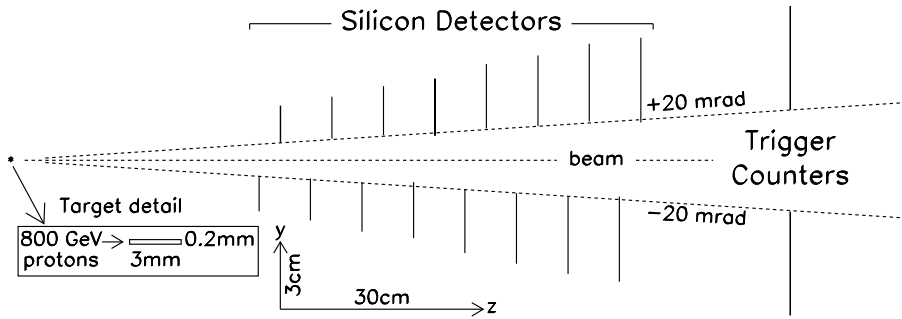}}
\vspace{-.05in}
\caption [E789  apparatus (plan view).]%
{E789 vertex telescopes (elevation view).}
\end{figure}

E789 results include the nuclear dependences of
charmonium production, which goes as $A^{\alpha(x_F)}$ with
$\alpha(0.05)=0.90\pm0.02$,~\cite{Kowitt,Leitch2}
and of $D^0 (\overline{D^0})$
production, which
depends linearly on target atomic weight
($\alpha=1.02\pm0.03\pm0.02$ at $x_F\simeq 0.03$).~\cite{Leitch}
The lower values of $\alpha$
observed for charmonium (especially for $x_F$ near zero)
are consistent with models in which charmonium
production is suppressed in nuclei due to dissociation by
comovers.~\cite{Leitch2} Models in which the nuclear suppression of charmonium
production is an initial-state effect (\eg due to possible nuclear modification
of the gluon structure function) are disfavored, since they would
predict similar nuclear dependences for $J/\psi$ and $D$ production.

We have measured differential cross sections for charm~\cite{Leitch} and
charmonium~\cite{Schub} production.
The observed $D^0$ cross section is
$d \sigma/dx_F = 58\pm3\pm7\,\mu$b/nucleon, which
extrapolated over all $x_F$ implies a total $D^0$ cross section $\sigma =
17.7\pm0.9\pm3.4\,\mu$b/nucleon).~\cite{Leitch}
Averaging with previous measurements using 800\,GeV proton
beams~\cite{Ammar_Kodama}  gives
$\sigma(pN\to D^0\,X)
+ \sigma(pN\to {\overline D {}^0} \,X) = (20.9\pm 3.5)\,\mu$b/nucleon,
consistent with next-to-leading-order (NLO) QCD predictions~\cite{Mangano}
 within the
broad range of theoretical uncertainty.
However, as at the Tevatron
Collider,~\cite{Braaten}
charmonium production is substantially underestimated, at least in models which
include only contributions from
color-singlet charmonium states below $D\overline D$
threshold.~\cite{Page}
We find $\sigma(p+N\to J/\psi+X) = 442\pm2\pm88\,$nb/nucleon and
$\sigma(p+N\to \psi^\prime+X) = 75\pm5\pm22\,$nb/nucleon, factors of 7 and 25
above QCD predictions.~\cite{Schub}

Production of beauty hadrons is studied by searching for evidence of
$J/\psi\to\mu^+\mu^-$ decay occurring in vacuum downstream of the 3-mm-long Au
target. A significant excess is observed of events with vertex downstream
of the target compared to those with vertex upstream, leading to
the measured cross section for $J/\psi$ from $b$ decay
$d^{\, 2}\sigma
/dx_F \, dp_T^2 = 107\pm28\pm19\,$pb/(GeV/$c$)$^2$/nucleon at $x_F = 0.05$
and $p_T = 1$\,GeV/$c$.~\cite{Jansen}
This can be corrected for the $b\to J/\psi+X$ branching ratio
and extrapolated over all of phase space to yield
$\sigma (p N \rightarrow b \overline{b} + X )
= 5.7  \pm 1.5 \pm 1.3\,$nb/nucleon.~\cite{Jansen}
This value is consistent with NLO QCD predictions
but a factor $\approx$2 below their central value.~\cite{Mangano}

The small beauty cross section at fixed-target energy
is surprising given that the beauty cross section at the Tevatron collider
is a factor 2 {\em above} the central QCD prediction.~\cite{Collider}
 It appears difficult to accomodate this steep $s$
dependence within the next-to-leading-order model.
However, caution is in order since the assumption of factorization of
$b$-quark production and fragmentation may be unreliable in
hadroproduction, especially at the low $p_T$ values of the fixed-target
data.~\cite{Mangano}

Considering that the range of QCD predictions included 3 nb,~\cite{Mangano} our
measurement
should be taken as favorable for the HERA-$B$ project, as is also our success
in reconstructing $b\to J/\psi+X$ decay vertices in a fixed-target environment
at a 50\,MHz interaction rate.
This latter accomplishment also indicates the feasibility of a fixed-target
charm experiment with sensitivity $\approx$2000 times higher than that of
existing experiments, for which a Letter of Intent is in
progress.~\cite{Charm2000}

\end{document}